\documentclass[usenatbib]{mn2e}
\usepackage{graphicx,ifthen,url}
\newcommand {\um}{$\mu$m}
\newcommand {\kms}{km\,s$^{-1}$}
\newcommand {\lya}{Ly$\alpha$}
\newcommand {\uJy}{$\mu$Jy}
\newcommand {\Mpy}{M$_{\sun}$ yr$^{-1}$}
\newcommand {\rga}{RG\,J123623} 
\newcommand {\rgb}{RG\,J123617} 

\newcommand {\aj}{AJ}
\newcommand {\apj}{ApJ}
\newcommand {\apjs}{ApJS}
\newcommand {\apjl}{ApJL}
\newcommand {\aap}{A~\&~A}
\newcommand {\mnras}{MNRAS}
\newcommand {\pasp}{PASP}
\newcommand {\physrep}{Phys. Rep.}
\newcommand {\nat}{Nature}
\newcommand {\araa}{ARA\&A}
\newcommand {\qjras}{QJRAS}

\begin{document}
\loadboldmathitalic
\title[Two z$\sim$2 MERLIN Evolved Radio Galaxies]{Constraining Star Formation and AGN in z$\sim$2 
Massive Galaxies using High Resolution MERLIN Radio Observations}
\author[C.M. Casey et al.]{C.~M. Casey$^{1}$\thanks{ccasey@ast.cam.ac.uk}, S.~C. Chapman$^{1}$, T.~W.~B. Muxlow$^{2}$, R.~J. Beswick$^{2}$, 
\newauthor  D.~M. Alexander$^{3}$, C.~J. Conselice$^{4}$ \\
$^{1}$ Institute of Astronomy, University of Cambridge, Madingley Road, Cambridge, CB3 0HA, UK\\
$^{2}$ Jodrell Bank Observatory, University of Manchester, Macclesfield, SK11 9DL, UK \\
$^{3}$ Department of Physics, University of Durham, South Road, Durham DH1 3LE, UK \\
$^{4}$ School of Physics and Astronomy, University of Nottingham, University Park, NG9 2RD, UK }

\date{Accepted 2009 February 6.}
\pagerange{\pageref{firstpage}--\pageref{lastpage}} \pubyear{2009}

\maketitle
\label{firstpage}

\begin{abstract}
We present high spatial resolution MERLIN 1.4GHz radio observations of two high 
redshift (z$\sim$2) sources, \rga\ (HDF147) and \rgb\ (HDF130), selected as the 
brightest radio sources from a sample of submillimetre-faint radio galaxies.  
They have starburst classifications from their rest-frame UV spectra.  However, 
their radio morphologies are remarkably compact ($<$80mas and $<$65mas respectively), 
demanding that the radio luminosity be dominated by Active Galactic Nuclei (AGN) 
rather than starbursts.  Near-IR imaging ($HST$ NICMOS F160W) shows large scale
sizes (R$_{1/2}\sim$0.75\arcsec, diameters $\sim$12kpc) and SED fitting to photometric 
points (optical through the mid-IR) reveals massive ($\sim$5$\times$10$^{11}$M$_{\sun}$), 
old (a few Gyr) stellar populations.  Both sources have low flux densities at 
observed 24\um and are undetected in observed 70\um\ and 850\um, suggesting a low 
mass of interstellar dust.  They are also formally undetected in the ultra-deep 2Ms 
$Chandra$ data, suggesting that any AGN activity is likely intrinsically weak.  We 
suggest both galaxies have evolved stellar populations, low star formation rates, 
and low accretion rates onto massive black holes (10$^{8.6}$M$_{\sun}$) whose radio 
luminosity is weakly beamed (by factors of a few).  A cluster-like environment has 
been identified near HDF130 by an over-density of galaxies at $z=1.99$, reinforcing 
the claim that clusters lead to more rapid evolution in galaxy populations.  These 
observations suggest that high-resolution radio (MERLIN) can be a superb diagnostic 
tool of AGN in the diverse galaxy populations at z$\sim$2.
\end{abstract}

\section{INTRODUCTION}

The development of deep, multiwavelength imaging surveys has shed light on the rapid construction
of galaxies and the build-up of stellar mass at high redshift (z$\sim$2-3).  The population of 
galaxies we see at this epoch is diverse$-$a testament to their rapid and significant growth.  
The cosmic star formation density peaks at this redshift (z$\sim$2) as does the quasar space density 
\citep{fan01a,blain02a,richards06a}.  Galaxies at z$\sim$2 display a wide range in properties, from 
negligible nuclear activity to the most powerful AGN, and with star formation rates (SFRs) over 1000 
M$_{\sun}$ yr$^{-1}$ in Submillimetre Galaxies \citep[SMGs; e.g.][]{blain02a} to as low 
as 1 M$_{\sun}$ yr$^{-1}$ \citep[e.g. quiescent galaxies; ][]{daddi05a,daddi04a,cimatti04a}.  Having also 
come to light in the past decade is the importance of AGN in regulating bulge growth \citep[e.g.][]{dimatteo05a}.  
This is apparent in the local M$_{\star}$-M$_{BH}$ relation where black hole mass is 
roughly $\sim$0.1$\%$ of the spheroid's stellar mass \citep{magorrian98a,kormendy95a,gebhardt00a}.  
This relationship clearly suggests that black hole mass and stellar mass build up symbiotically, with 
a broad range of z$\sim$2 galaxies lying within a factor of a few of the local M$_{\star}$-M$_{BH}$ 
ratio \citep[e.g.,][]{peng06a,mclure06a,alexander08a}.  Moderate-luminosity AGN with apparently low accretion 
efficiencies, ultra-luminous AGN with powerful feedback winds, and heavily obscured, Compton thick AGN 
whose strength is shielded by dust in the host galaxy have been identified in galaxies with L$_{X}\ >$ 
10$^{43}$ erg s$^{-1}$ at high-$z$ \citep{brandt05a,alexander05a,alexander08b,daddi07a,comastri04a}.  
Only through the most systematic searches of deep survey data may the full range of z$\sim$2 galaxy 
properties be clearly understood.

In an effort to study an active subset of z$\sim$2 galaxies, our collaboration has focused on the 
\uJy\ radio galaxies, which are roughly 1/3 AGN, 1/3 submillimetre-bright star formers (SMGs) and
1/3 submillimetre-faint star formers (submillimetre-faint star forming radio galaxies, SFRGs).
While most SMGs and SFRGs may have some contribution from low-luminosity AGN, their bolometric
luminosity is dominated by dust-generated emission in the far-IR.  The AGN sources have bolometric
luminosities dominated by non-thermal radiation from AGN.  SFRGs are likely to be the hotter dust 
counterparts of the SMG population \citep{chapman04a}, the difference in their submillimetre properties 
caused by a shift in the blackbody dust emission towards higher temperatures (T$_d>$50\,K).  There is no 
other obvious physical difference between SFRGs and SMGs, both contributing significantly to the z$\sim$2 
star formation history (Casey et al. 2008, in preparation).  Here we present detailed multiwavelength 
studies of two galaxies, \rga\ (HDF147) and \rgb\ (HDF130), originally identified as SFRGs but 
postulated here as being radio-bright but intrinsically low-luminosity AGN.

In \S \ref{sampleobs_s} of this paper we describe the observations and the parent sample from which 
the two galaxies are drawn.  The observations include the MERLIN radio survey, VLA coverage of HDF, 
coverage by $Spitzer$ 24\um, 70\um, $HST$ photometry and morphology, and X-Ray from CDF-N. In \S 
\ref{results_s} we interpret these observations and derive an SED model, while in \S \ref{discussion_s} 
we discuss the selection population and the possible physical scenarios that could describe such a set 
of multiwavelength observations, from a massive yet poorly accreting black hole whose radio emission 
is weakly beamed, to strongly beamed low-luminosity AGN, and to heavily obscured Compton thick AGN.
Throughout we assume H$_0$=70\kms, $\Omega_0$=0.3 and $\Omega_\Lambda$=0.7.

\section{SAMPLE AND OBSERVATIONS}\label{sampleobs_s}

\begin{table}
\begin{center}
\caption{Physical Properties of HDF147 and HDF130} 
\label{table1}
\begin{tabular}{lrr}
\hline
    &   HDF147   &  HDF130 \\
\hline 

RA (J2000)           &         12:36:23.54 & 12:36:17.55  \\
DEC (J2000)          &   +62:16:42.8       & +62:15:40.8  \\
redshift             &               1.918   & 1.993 \\
VLA S$_{\rm 1.4GHz}$      & 481.0$\pm$25.4\,\uJy        & 200.0$\pm$12.8\,\uJy \\
WSRT S$_{\rm 1.4GHz}$     & 476$\pm$31\,\uJy        & 375$\pm$28\,\uJy \\
S$_{850\mu m}$    & 1.6$\pm$1.1\,mJy           & 2.1$\pm$1.0\,mJy \\
S$_{70\mu m}$       & $<$2.1\,mJy                & $<$1.9\,mJy \\
S$_{24\mu m}$   & 18$\pm$3\,\uJy              & 24$\pm$3\,\uJy \\
8.0\um\ mag           & 20.45$\pm$0.59      & 21.48$\pm$0.95 \\
5.8\um\ mag           & 19.76$\pm$0.74      & 20.77$\pm$1.17 \\
4.5\um\ mag           & 19.57$\pm$0.32      & 20.58$\pm$0.52 \\
3.6\um\ mag           & 19.63$\pm$0.31      & 20.72$\pm$0.52 \\
B mag              & $>$25.7              & 25.4$\pm$1.0 \\
V mag              & $>$26.5              & 25.3$\pm$1.0 \\
I mag              & 25.9$\pm$1.0         & 25.1$\pm$1.0 \\
Z mag              & 24.6$\pm$0.5         & 24.7$\pm$0.5 \\
H mag              & 21.2$\pm$0.5         & 22.5$\pm$0.6 \\
L$_{\rm X,(0.5-8.0keV)}$ & 4.1$\times$10$^{42}\,erg\,s^{-2}$    & $<$8.2$\times$10$^{42}\,erg\,s^{-2}$  \\
$R_{1/2}$ (H band) & 8.65kpc (1.02\arcsec)          & 3.7kpc (0.44\arcsec) \\
$R_{1/2}$ (MERLIN) & $<$680pc ($<$80mas)             & $<$550pc ($<$65mas)  \\
log(M$_\star$/M$_{\sun}$)  &  11.8$^{+0.3}_{-0.3}$                & 11.4$^{+0.4}_{-0.2}$  \\
log(M$_{BH}$/M$_{\sun}$)  &  8.9$^{+0.6}_{-0.6}$                & 8.5$^{+0.7}_{-0.5}$  \\
$\rho_{\star}$ (M$_{\sun}$\,kpc$^{-3}$) &   1.2$^{+1.1}_{-0.6}\times$10$^{8}$  &  5.7$^{+8.7}_{-2.1}\times$10$^{8}$ \\
n (Sersic Index)    &   3.5$\pm$0.3              &  1.5$\pm$0.5  \\
H-band axis ratio      &  0.66              &  0.98  \\
$E(B-V)$            &    0.40        & 0.03 \\
Ly$\alpha$/C\,IV    &  0.75$\pm$0.06  & 1.67$\pm$0.03  \\
Ly$\alpha$/He\,II    & 0.62$\pm$0.06  & 1.67$\pm$0.02  \\
\hline
\end{tabular}
\end{center}
Notes $-$ All magnitudes are in the AB-magnitude system.
X-Ray luminosity is given for the rest frame 0.5-8.0keV band.
The variable $R_{1/2}$ denotes the half light radius of the galaxy in 
H band and MERLIN radio imaging respectively.  The Sersic Exponent, n,
refers to the {\sc Galfit} profile fitting to H-band.  The derived 
quantities come from calculations described in \S \ref{results_s}.
\end{table}

HDF147 and HDF130 (see Table~1 for observation details) originate from a sample of 18 SFRGs
\citep[also referred to as `OFRGs';][]{chapman04a}.  The sample was defined by all 
radio-detected galaxies in seven distinct fields that were previously targeted in the sub-mm by SCUBA 
\citep{holland99a} but lacked an optical counterpart brighter than R = 23.5.  As described in \citet{chapman04a}, 
the requirement of an optically faint source eliminates low redshift, moderate-luminosity sources as well as high 
redshift, bright AGN and leaves a proposed population of predominantly hot dusty ultra-luminous galaxies at z 
$\approx$ 2.  The set of SFRGs have much lower submillimetre flux densities ($\sim$0.5 mJy) than the nominal 
$\sim$6.5 mJy of SMGs \citep[e.g.][]{chapman05a}.

High-resolution observations from the Multi-Element Radio Linked Interferometer Network 
\citep[MERLIN;][]{thomasson86a} were obtained for these sources as described in \citet{muxlow05a}.
One motivation for the high angular resolution observations is distinguishing the 
compact emission of an AGN from more diffuse star formation, separating star forming galaxies from AGN.  
\citet{muxlow05a} discusses the 0.3-0.5\arcsec\ beamsize MERLIN observations in depth.
The rms noise from MERLIN data was 5.9 \uJy\ beam$^{-1}$.  The images from the VLA \citep{richards99a,
  richards00a}  and MERLIN were combined in the image plane and used to make a sensitive 1.4-GHz map 
with high positional accuracy (tens of mas) allowing high resolution imaging and identification 
of \uJy\ sources.  

$Spitzer$ observations were made in IRAC 3.6\um, 4.5\um, 5.8\um\ and 8.0\um\ and MIPS 70\um\ \citep{frayer06a} 
and 24\um\ bands in the HDF region (PI M. Dickinson; R. Chary et al., in prep). This component of observations 
is used to study the galaxies' stellar masses. Optical Imaging of the area is from the Hubble Deep Field\footnote{
Based on observations made with the NASA/ESA Hubble Space Telescope, and obtained from the Hubble Legacy Archive, 
which is a collaboration between the Space Telescope Science Institute (STScI/NASA), the Space Telescope European 
Coordinating Facility (STECF/ESA) and the Canadian Astronomy Data Centre (CADC/NRC/CSA).}, 
in B (F435W), V (F606W), i (F814W), and z (F850LP) bands all in the AB magnitude system.  As part of the 
extensive 45\,arcmin$^2$ near infrared survey coinciding with the HDF and GOODS-N, we have obtained NICMOS 
imaging using NIC3 three orbit pointing in F160W of both HDF147 and HDF130 (Conselice et~al., in prep).

The parent sample of SFRGs was also observed spectroscopically with Keck/LRIS \citep{steidel04a,oke95a} 
from 2002-2004.  Full details of these rest-UV spectral observations are found in \citet{chapman04a}.  
In short, the galaxies' redshifts were identified by detection of several emission features (\lya\ and C\,IV 
in the case of HDF147 and \lya\ and He\,II in the case of HDF130).  Although their spectra are faint (see 
Figure~2 of Chapman et~al.), both resemble starburst galaxies, so we grouped 
them originally in the parent SFRG sample.  We attempted to measure individual line fluxes from several 
features that would more clearly classify HDF147/HDF130 as starbursts or AGN, but the spectra are too faint 
to go beyond the analysis of \citet{chapman04a}.

Observations in the X-Ray are from the 2 Ms $Chandra$ Deep Field North 
\citep[CDF-N;][]{alexander03a}, centered on the HDF, in the 0.5-8.0 keV 
band which achieves an aim-point sensitivity of $\approx$1.2 $\times$ 10$^{-16}$ ergs cm$^{-2}$ 
s$^{-1}$.  X-Ray observational details of a similar SMG population are 
found in \citet{alexander05a}.  At the redshifts of these galaxies 
($\sim$1.96), the observed 0.5-8.0\,keV band corresponds to rest frame 
energies 1.5 - 24\,keV.  Neither galaxy is listed in the \citet{alexander03a}
catalog; however, HDF147 is marginally detected at a low significance in the narrow 1-2\,keV
band (a luminosity of 1.67$\times$10$^{42}$ erg s$^{-1}$ in the observed 0.5-8\,keV band).
Assuming an average X-ray spectral slope 
$\Gamma$ = 1.8, we find the luminosity of HDF147 in the rest 0.5-8.0\,keV band is 
L$_{X (HDF147)}$ = 4.1$\times$10$^{42}$ erg s$^{-1}$ and the the 3-$\sigma$ upper 
limit for HDF130 is L$_{X (HDF130)}$(2-10\,keV)$\le$8.2$\times$10$^{42}$ erg s$^{-1}$.

\section{Analysis and Results}\label{results_s}

\subsection{Constraining Star Formation}

The measurement of the radio angular sizes of HDF147 and HDF130 were made using the {\sc AIPS} software,
{\sc JMFIT}.  {\sc JMFIT} was used to fit a single 2-d Gaussian to the core component of each galaxy
in the CLEANed MERLIN plus VLA radio image restored with a 200mas circular beam.  Subsequently the 
restoring beam of this image was deconvolved from these fitted sizes.  The radio images of both HDF130 
and HDF147 are of high signal-to-noise and close to the pointing centre, thus the measurement of 
deconvolved radio sizes several times smaller than the restored CLEAN beam of the image can be reliably 
made \citep[e.g.][]{condon98a}.  The fitted sizes, deconvolved with the image restoring beam, of the 
cores of HDF147 and HDF130 are $<$80mas and $<$65mas respectively, with an error on the fitted FWHM of 
20mas \citep{condon97a}.  At these redshifts, these angular size limits correspond to linear sizes of 
$<$640pc and $<$520pc.

Figure~1 shows contoured images of HDF147 and HDF130 overlaid on ACS/NICMOS imaging.  Muxlow et~al. (2005)
classify HDF147 (J123623+621642) as an AGN candidate and note that it has a faint one-sided radio emission 
extending $\sim$0.6\arcsec\ to the south of the compact core.  This core and jet-like emission can clearly 
be seen in Figure~1.  The radio structure of HDF130 (J123617+621540) is compact in the high resolution 
images presented here, however from lower resolution (0.5\arcsec) imaging of these data Muxlow et~al. 
(2005) showed this source to have compact core with two-sided radio emission extending over a linear extent 
of 1.3\arcsec.  The HDF147 radio morphology of Fig. 1 exhibits a low luminosity ($\sim$20\uJy) extension 
along the galaxy's minor axis.  This is a possible candidate for radio jet outflows but provides no obvious 
clue to source orientation.  

If star formation were the source of radio emission, both sources would exceed the maximum star 
formation density possible for a given mass and dynamical timescale \citep{elmegreen99a}. HDF147 
and HDF130 have limits on their radio size smaller than most local ULIRGs.  As local 
ULIRGs with $\Sigma_{SFR}\,\sim$200\Mpy kpc$^{-2}$ are already forming stars near a theoretical maximum 
\citep[e.g.,][]{tacconi06a}, HDF147 and HDF130 would exceed this limit by factors of 50-70$\times$ 
given that their radio-inferred SFR densities are 15000\Mpy kpc$^{-2}$ and 11000\Mpy kpc$^{-2}$
respectively (with only small variations due to geometry and gas fractions).  We therefore conclude 
that the radio emission we observe must be largely emitted by AGN.  

We further constrain their star formation rates using the rest-UV luminosities 
\citep[Equation~1 of ][]{kennicutt98a}.  Without removing the effects 
of dust extinction, we constrain HDF147 to a maximum\footnote{If all 
of the UV emission were from star formation.} SFR of 5$^{+4}_{-2}$\,\Mpy\ 
and HDF130 to an SFR of 17$^{+6}_{-4}$\,\Mpy.  Corrected for extinctions 
of E(B-V)=0.40 and 0.03 respectively\footnote{We considered the effect of 
IGM extinction at z$\sim$2, but found it was insignificant compared to
magnitude uncertainty and extinction internal to the galaxies.} (calculated from {\sc HYPERZ} 
stellar population model fits, as described in the next section), the 
UV SFR constraints are SFR$_{HDF147}$=110$^{+88}_{-44}$\,\Mpy\ 
and SFR$_{HDF130}$=22$^{+8}_{-6}$\,\Mpy.  The low 24\um\ and 850\um\ luminosities
are consistent with these SFRs (S$_{850}\le\,2\,mJy$; $SFR\,<\,$300\Mpy).

\subsection{Characterizing the Stellar Populations}\label{seds_s}

To learn about the stellar components of HDF147 and HDF130, we fit the optical and 
$Spitzer-IRAC$ photometric points to stellar population templates from \citet{maraston05a}
using the {\sc hyperz} package \citep{bolzonella00a}.  {\sc hyperz} output includes internal
extinction factor $A_V$ and rest-$K$ band absolute magnitude, which we use to calculate
extinction-corrected UV SFRs and derive stellar masses respectively.  The Maraston model 
populations assume a Salpeter IMF and range in age from 10\,Myr to 15\,Gyr\footnote{Although 
the age of the Universe is only $\sim$13.7\,Gyrs, the aging of the Maraston stellar population 
models has not been normalized to the current cosmology, thus allowing ``15 Gyr'' old models.}.
We find that the older models (10-15\,Gyrs) are much better fits to both galaxies' photometry
than a `young' model aged only a few Gyrs.  Furthermore, we find that the instantaneous starburst
models with 0.67 metallicity fit best, where goodness of fit is measured by minimizing the
$\chi^{2}$ statistic.

\begin{figure}
\begin{center}
\includegraphics[width=0.99\hsize]{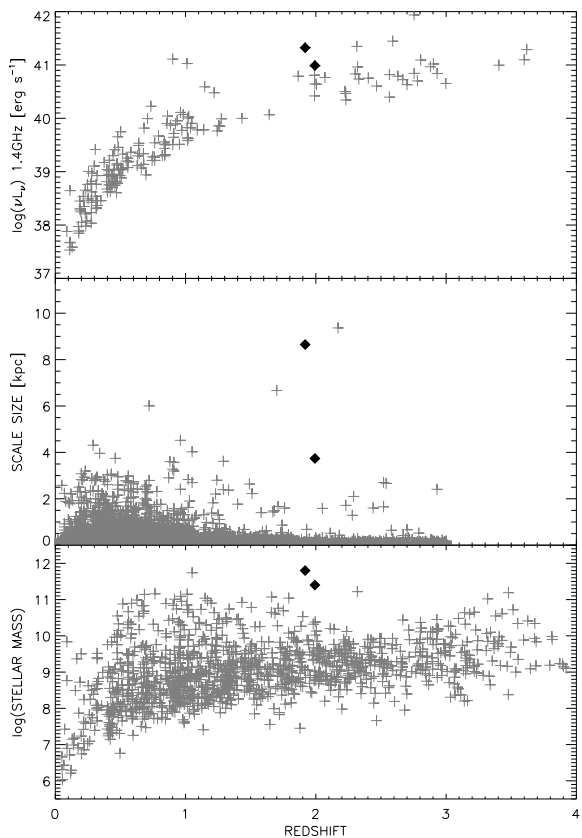}
\caption{ 
  Radio luminosity \citep{chapman03c}, scale sizes \citep{abraham03a,abraham04a}, and 
  stellar mass \citep{conselice08a,conselice05a} against redshift for radio galaxies 
  in several large survey fields.  HDF130 and HDF147 are highlighted as black diamonds 
  at much larger radio luminosities, scale sizes, and higher masses than are typical 
  high-z galaxies.
}
\label{selection_f}
\end{center}
\end{figure}

We measure extinction factors $A_V=1.3$ for HDF147 and $A_V=0.1$ for HDF130, which corresponds
to $E(B-V)_{147}=0.40$ and $E(B-V)_{130}=0.03$.  These derived values are used to estimate the
rest-UV star formation rates described in the previous section. 

We have used the methodology outlined in \citet{borys05a} to estimate stellar masses from rest 
frame $K$-band magnitudes, which we measured from {\sc hyperz} to be -26.67 for HDF147 and -26.81
for HDF130.  We assume a mass to light ratio of L$_K$/M = 1.9 because of the galaxies' predominantly 
old stellar populations ($\tau$$>$1\,Gyr).  Both galaxies are exceptionally massive, with log(M/M$_{\sun}$) 
= 11.8$^{+0.3}_{-0.3}$ and 11.4$^{+0.4}_{-0.2}$ for HDF147 and HDF130 respectively.  The contrast to other 
z$\sim$2 galaxies is seen in panel 3 of Figure \ref{selection_f}.  We use the well studied local 
M$_{\star}$-M$_{BH}$ relation where M$_{BH}$ = M$_\star/1000$ \citep{magorrian98a,kormendy95a,gebhardt00a} 
to estimate black hole masses of log(M$_{BH}$/M$_{\sun}$) = 8.9$^{+0.6}_{-0.6}$ and 8.5$^{+0.7}_{-0.5}$, with 
the caveat that there is a factor of two variation in the M$_{BH}$/M$_{\star}$ ratio.  We caution that 
M$_{\star}$-M$_{BH}$ has a larger spread and is slightly elevated at z$\sim$2 than locally 
\citep[e.g.,][]{peng06a,mclure06a} although such variation could be caused by a bias towards large black 
holes \citep{lauer07a,alexander08a}.

\begin{figure*}
\includegraphics[width=1.8\columnwidth]{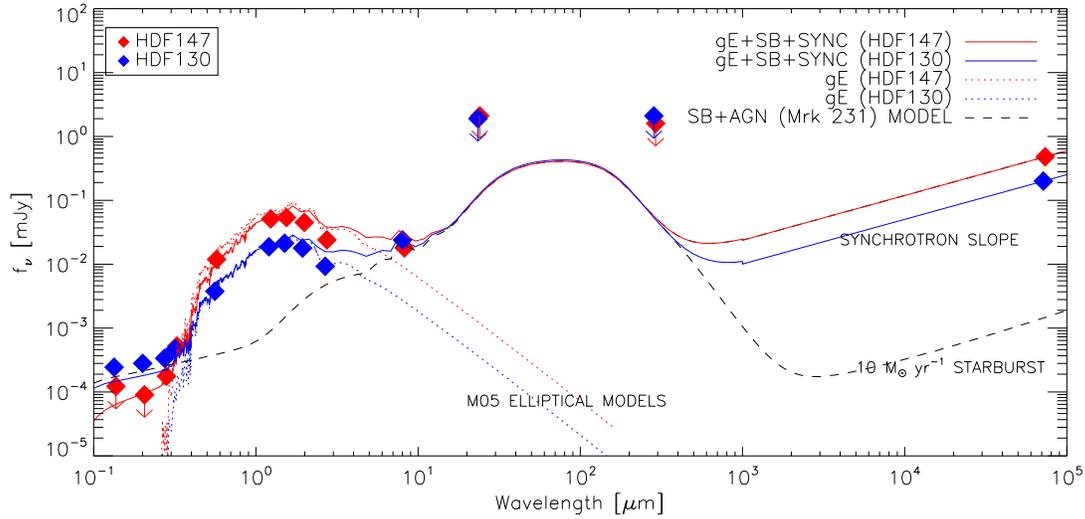}
\begin{center}
\caption{
  A composite SED for the two galaxies (red for HDF147, blue for HDF130) in rest wavelength, 
  from UV through the infrared, submillimetre, and radio.  The old stellar populations, 
  adapted from \citet{maraston05a} stellar evolution models, are fit to the optical and 
  $Spitzer-IRAC$ photometry, around the rest-1.6\um\ stellar bump (dotted lines labeled 
  `M05 Elliptical Models').  A Calzetti dust extinction law \citep[e.g.][]{calzetti94a} is used to extinct the flux 
  in the UV, with derived reddening factors of $E(B-V)=0.40$ and $E(B-V)=0.03$ for HDF147 
  and HDF130 respectively.  A Mrk\,231 template SED \citep{berta06a}, scaled to the 24\um\ 
  flux densities, is added to represent the far-IR and radio portion of the SED.  We fit 
  the SED to the radio points by scaling up the emission from synchrotron emission 
  ($\alpha\,=\,0.7$) from the nominal flux densities that were expected from the Mrk\,231 
  template (dashed line).  
}
\label{sed_f}
\end{center}
\end{figure*}

Figure \ref{sed_f} explores SED fitting for these galaxies from the rest-UV wavelengths to the radio.  
The SED shortward of 1\um\ is dominated by the best-fit {\sc hyperz} Maraston stellar population models (dotted
lines in Fig.~\ref{sed_f}).  Towards longer wavelengths, we introduce a template Mrk\,231 SED \citep[adapted 
from][]{berta06a} including both AGN and moderate star former components and normalize it to the 24\um\ flux 
densities (dashed line in Fig.~\ref{sed_f}).  The inferred star formation rate from templates with this 24\um\ 
normalization is $\sim$10\Mpy.  We then raise the radio synchrotron contribution to the SED (but we keep the synchrotron
slope, $\alpha=0.7$ the same) to fit the galaxies' bright radio luminosities.

The optical to mid-IR portion of the SED provides a good fit to the data: a sum of old and massive 
stellar populations with moderate ($\sim$10\,M$_\odot$ yr $^{-1}$) star formation.  The submillimetre flux 
density limits at do not place any significant additional constraints on the SEDs.  In the rest-UV ($B$ 
\& $V$ bands) HDF147 is un-detected.  HDF130 exhibits a moderate blue excess, with associated SFR=22\Mpy, 
which may correspond to the starburst detected in its rest-UV spectrum.  We emphasize that the total 
luminosity from the star forming component is minor compared to the very luminous SED-characterized 
``1.6\um-bump'' stellar population.

\subsection{Size and Morphological Classification}

\begin{figure*}
\begin{center}
\includegraphics[width=1.99\columnwidth]{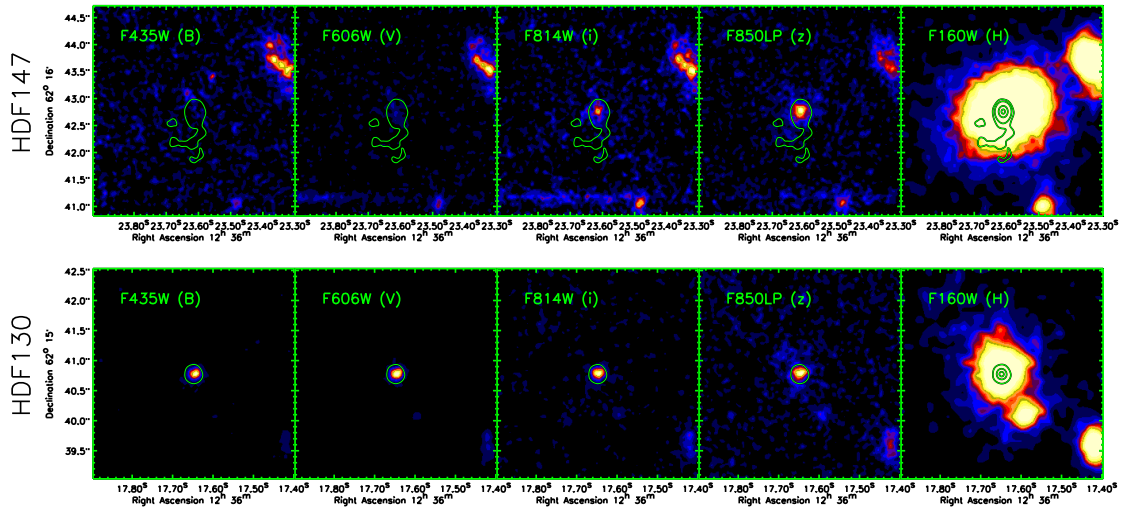}
\caption{ MERLIN radio contours overlayed with ACS B, V, i, z (F435W, F606W, F814W, F850LP) 
  and NICMOS H (F160W) imaging of galaxies HDF147 (top) and HDF130 (bottom).  HDF147 shows
  a faint extended region to the north of the radio and NICMOS centers; its NICMOS size is
  substantially larger than most galaxies at its epoch at a $\sim$17\,kpc diameter.  HDF130 
  has compact blue emission which is coincident with the radio center, but has a much more 
  extended morphology, $\sim$8\,kpc across, in $H$-band.  The ACS panels show only the 15\uJy\ 
  radio contour for clarity while the NICMOS panels show more complete contours: 15\uJy, 
  40\uJy, 150\uJy, 255\uJy, and 380\uJy.  The peak flux density of HDF147 is 423\uJy\ and 
  for HDF130 it is 162\uJy.
}
\label{radio2d_f}
\end{center}
\end{figure*}

In contrast to optical imaging, NICMOS near-IR imaging reveals that the galaxies have 
 half light radii of 8\,kpc and 4\,kpc, over 20 times the scale of the 
radio core emission.  This is shown in Figure \ref{radio2d_f}.  Van Dokkum et al. (2008) 
find that while local galaxies have a typical size of R$_{1/2}\sim$5\,kpc, elliptical galaxies 
of $\sim$3$\times$10$^{11}$\,M$_{\sun}$ at z$\sim$2.3 are much more compact with average 
R$_{1/2}$=0.9\,kpc, which is significantly smaller than HDF147 and HDF130 (see Figure 
\ref{selection_f}, panel 2).  However the stellar densities of HDF147 and HDF130 
($\rho_{\star,147}$=1.2$^{+1.1}_{-0.6}$\,M$_{\sun}$\,kpc$^{-3}$, $\rho_{\star,130}$=5.7$^{+8.7}_{-2.1}$\,M$_{\sun}$\,kpc$^{-3}$
\footnote{The mean stellar density within the half light radius, 
$\rho_{\star}=0.5M/(\frac{4}{3}\pi R_{1/2}^{3})$.}) are much lower than the high-density
z$\sim$2.3 galaxies, and similar to local ellipticals \citep{blanton05a}.
A similar size comparison of massive (M $>$ 10$^{11}$M$_{\sun}$) 
high-z (1.7$<$z$<$2.0) galaxies from \citet{trujillo07a} also shows that HDF147 and HDF130 are much 
larger than typical quiescent galaxies at similar redshifts.
\citet{trujillo07a} find that for galaxies with large Sersic index (n$>$2.5) the expected size is 
$R_{1/2}$=1.1$\pm$0.4\,kpc and for shallower potentials (n$<$2.5) it is $R_{1/2}$=2.3$\pm$1.8\,kpc.

We fit Sersic profiles to H-band imaging to search for substructure, derive Sersic indices, 
and model their morphologies. Fitting both HDF147 and HDF130 using {\sc Galfit} 
\citep{peng02a}, we find that HDF147 closely resembles a de Vaucouleurs profile 
(n=3.5$\pm$0.3) while HDF130 fits well to a more extended exponential (n=1.5$\pm$0.5).  
Clearly by this analysis, HDF147 is a giant elliptical galaxy with smooth morphology 
and R$_e\sim$9\,kpc.  HDF130 is near the boundary between elliptical galaxies and disk-like 
structures at n=1.5 and shows no substructure; its giant size, giant mass, lack of significant 
star formation ($<$100\Mpy), and old stellar population are characteristic of large quiescent galaxies.

\subsection{Environment Characterization and Further Classification}

Given the large stellar masses, old stellar populations, and large scale sizes of 
these galaxies, the question of their environment naturally arises since locally giant elliptical (gE) 
galaxies reside exclusively in highly clustered environments.  \citet{blain04a} observed that there 
was a strong association of 5 SMGs in this field at the redshift $z\sim1.99$ (around HDF130), representing 
the largest {\it cluster} of SMGs in the \citet{chapman05a} spectroscopic survey. \citet{chapman08a} have 
studied this structure further showing that an over-density of galaxies of all types exists at this redshift, 
although not with the same degree of contrast as the SMGs. No such structure was found associated with HDF147,
although it is possible that one exists and has not been observed due to spectroscopic incompleteness.

The \citet{bauer02a} study of extended X-ray sources detects elongated X-ray emission roughly 
centered on HDF130 (CDF-N source 2, scale size $\sim$1\arcmin, see their Figure~2 most upper 
right panel).  Due to its double lobed shape and sharp cutoff on the ends,  the emission could
be due to inverse Compton scattering of the CMB due to AGN outflow in HDF130 rather than 
the cooling gas of an extended cluster.  The size of the emission region is 300-500\,kpc, and 
no radio lobes or plumes are observed to correlate with the X-ray region.  Detailed analysis 
of the X-ray observations of this region are presented in Fabian et~al. (2009, in prep).

These galaxies were not selected as $z=2$ BX/BM objects by \citet{steidel04a} from their colors, 
and they do not appear in the spectroscopic catalogs of \citet{cowie04a}. Both galaxies satisfy 
the {\it p-BzK} selection criteria for passively evolving galaxies ($BzK\ <$ -0.2, $z-K\ >$ 2.5)
as described by \citet{daddi04a}.  However, the scale size of these galaxies is much larger than
the vast majority of p-BzK galaxies which are generally very compact \citep[scale radii 
$<$0.9\,kpc;][]{vandokkum08a}.

\section{DISCUSSION}\label{discussion_s}

\begin{figure}
\begin{center}
\includegraphics[width=0.99\hsize]{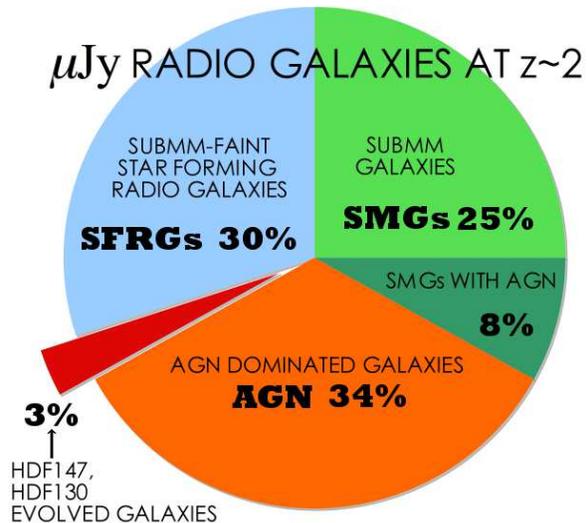}
\caption{ 
  A schematic pie chart characterizing the z$\sim$2 \uJy\ radio source population.  SMGs and SFRGs have optical
  spectra consistent with starbursts, while AGN and SMGs with AGN have spectral signatures of AGN.  The SMGs and
  SMGs with AGN are detected at 850\um.  HDF147 and HDF130 are highlighted as the 
  small red slice, between SFRGs and AGN; while they were previously diagnosed as SFRGs from their spectra, we
  have shown that they are more appropriately classified as AGN.
}
\label{piechart_f}
\end{center}
\end{figure}

To understand the importance of galaxies like HDF147 and HDF130 at z$\sim$2, we first 
characterize the parent sample of radio sources from which they were selected.
We illustrate schematically in Figure \ref{piechart_f} the relative fractions of 
different galaxies (SMGs, SFRGs and AGN) comprising the \uJy\ z$\sim$2 radio galaxy 
population (a more detailed analysis is in Casey et al. 2008, in preparation).  HDF147 
and HDF130 are highlighted since our observations have demonstrated that they should be 
classified as AGN-dominated rather than star formation-dominated SFRGs.  These two 
galaxies represent a rare type of massive galaxy with an estimated volume density of 
1.5$\times$10$^{-6}$\,Mpc$^{-3}$ (given two objects detected in a well studied area 
15\arcmin$\times$15\arcmin\ and redshift range 1.8$<$z$<$2.6).  Evolved elliptical 
galaxies of this type (with stellar masses over 10$^{11}$ M$_{\sun}$ and negligible 
black hole accretion) are extremely rare at high redshift.  \citet{conselice07a} 
estimate $\Phi$ $\ll$ 5$\times$10$^{-6}$ h$_{70}^{3}$ Mpc$^{-3}$ the volume density estimate 
for highly evolved z$\sim$2 galaxies, which is less than $\Phi$ for SMGs \citep{chapman03c}.   
While this suggests a quite sparse population, a combination 
of spectroscopic incompleteness in the radio source population and a possible wider 
range of physical properties consistent with these two galaxies could suggest that 
they are an important evolutionary stage of massive galaxies.

This section explores physical explanations for the observational constraints: high-resolution radio
indicating the presence of an AGN, low X-Ray luminosities, no evidence for high ionization UV emission lines 
and low 24\um\ luminosities.

\subsection{Evolved Galaxies with Poorly Accreting Black Holes?}

The observational constraints (low 24\um\ and X-Ray luminosities) require that there be very little 
accretion ($\eta \ll $0.1) onto the massive 10$^{8.9}$ and 10$^{8.5}$ M$_{\sun}$ black holes.  This 
lower accretion efficiency could be explained by decreased fueling of the AGN, where most cool gas 
has been converted into stars or expelled from the elliptical galaxies by feedback mechanisms.  A 
lack of hot AGN torus dust is a condition for this interpretation because of the weak mid-IR luminosity.

With low X-Ray luminosities yet high radio luminosities, we suggest these radio fluxes may be boosted by some 
mechanism.  We propose weak radio beaming, here defined as jet-line of sight orientation angles less than 30$^{o}$ with 
fluxes boosted by a factor of a few, to explain the 200-500\uJy\ radio flux densities.  At the other extreme, 
referring back to Figure \ref{sed_f}, we see that the radio emission inferred from a moderate-luminosity 
AGN with modest star formation would be $\sim$2\uJy.  The intrinsic emission of the HDF147/HDF130 AGN components 
must lie between 2 and 200\uJy.  

We suggest that HDF147 and HDF130 are likely the high-redshift analog of local Fanaroff-Riley\,I's \citep{fanaroff74a}, 
which we use to estimate volume density and the probability of our observations.  Both galaxies fall below the 
critical FR\,I/FR\,II boundary in radio luminosity and rest-frame optical magnitude \citep[i.e. the radio-optical 
magnitude bivariate FR\,I/FR\,II boundary;][]{ledlow96a}.  They are also classified as FR\,I's in the original
classification scheme \citep[i.e. $L\,<\,$L$_{\rm CRIT}(178\,MHz)\,\approx\,$2$\times$10$^{25}$\,W\,Hz$^{-1}$\,str$^{-1}$ assuming
$\alpha=0.8$;][]{fanaroff74a}.  Their intrinsic emission is 
likely similar to the average luminosity of these nearby FR\,I AGN in evolved galaxies with $\nu$L$_{\nu}(1.4GHz)\sim$5$\times$10$^{39}$\,erg\,s$^{-1}$.
The probability of observing intrinsic radio luminosities of $\nu$L$_{\nu}$ (1.4\,GHz) $\sim$ 10$^{41.3}$ erg s$^{-1}$
among FR\,I radio sources is very low \citep{ledlow96a}: 0.5\% (to observe HDF147 at $\nu$L$_{\nu}$ = 10$^{41.5}$)
and 1.3\% (to observe HDF130 at $\nu$L$_{\nu}$ = 10$^{41.1}$).  Both probabilities represent the fraction of FR\,Is in 
the Ledlow \&\ Owen sample with that luminosity or greater.  If we allow beaming in both cases, so that intrinsic 
luminosities drop by a factor of ten, then the probability of our observation increases marginally, by 5 times.  We 
caution that the statistics behind the \citet{ledlow96a} sample are limited and only include nearby sources, and 
that the evolution with redshift might not be constant.

We can also use the much larger and well studied statistics
of the faint-end quasar luminosity function, and apply constraints based on different beaming-strength geometries.
The most recent studies of the quasar luminosity function (QLF) are from \citet{hopkins07a} and \citet{richards06a}, 
whose faint-end reliabilities, as they relate to probable low-luminosity low-accretion rate AGN, are discussed in 
\citet{casey08a}.  For 26.5$<$i$<$23.5 (the magnitude selection cut on the parent sample of SFRGs), and 1.7$<$z$<$2.3, 
the AGN surface density is 168$\pm$2 deg$^{-2}$.  Assuming an isotropic distribution in orientation angles for
type\,1 AGN, the number of these faint AGN whose jets are pointed towards Earth within 30$^{o}$ (weak-beaming) 
is 42$\pm$17 deg$^{-2}$.  For strong beaming ($<$3$^o$ opening angle, $\sim$100 amplification) the numbers shrink to 
0.8$\pm$0.3 deg$^{-2}$.  We therefore expect 1.2$\pm$0.5 low-luminosity, weakly beamed AGN to 
appear in the MERLIN coverage of GOODS-N, which agrees with the two galaxies we have found.

Since we classify these galaxies as massive, with old stellar populations, there is also a constraint
on X-Ray emission produced by hot gas that we test.  While all massive ellipticals in the local Universe
have large quantities of X-Ray emitting gas \citep[as have been shown out to $z=0.7$ in][]{lehmer07a}, the 
expected X-Ray luminosity from hot gas in these systems is about an order of magnitude below the current 
$Chandra$ X-Ray limits (L$_X$$\sim$10$^{41}$\,erg\,s$^{-1}$)\footnote{The hot-gas X-Ray constraints were 
calculated using the average X-Ray to K-band luminosity ratio of early type galaxies from \citet{ellis06a}, 
\citet{lehmer07a}, and \citet{lehmer08a}.}. 

\subsection{Alternate Hypotheses}

The possibility exists that HDF147/HDF130 are highly-beamed (opening angle $<$3$^{o}$)
low-luminosity AGN.   This hypothesis would further reduce the concern for low dust 
masses in a hot torus since low-luminosity AGN would not create significant re-radiated 
mid-IR flux even in the presence of dense dust clouds.  However, looking straight down 
the radio jets, one would expect signs of a BL Lac type optical spectrum, which is 
clearly not present.  A variation on the low-luminosity AGN hypothesis is that the 
central black holes are much less massive than the local M$_{\star}$-M$_{BH}$ predicts.  
The lower mass limits determined by Eddington limited accretion are 10$^{5.9}$ and 
$<$10$^{6.1}$ M$_{\sun}$ \citep[two orders of magnitude below the predictions of M$_{\star}$-M$_{BH}$, 
as in Figure 3 of][]{borys05a}.  However, under this hypothesis we would have to advocate 
factors $>$100$\times$ beaming to boost the radio luminosity above what is expected for 
such small black holes \citep[e.g.,][]{merloni03a}.  With the analysis of the previous 
subsection, the orientation geometry is likely implausible for either variation of the 
highly-beamed, low-luminosity AGN hypothesis (with probability $\ll$1\%, requiring both 
galaxies to have jet-line of sight orientation angles of $<$3$^{o}$).

Compton thick AGN are also an alternate hypothesis.  The lack of detection in the 
soft X-ray could be explained by a deeply buried, obscured AGN.  The extreme obscuration could also
explain the weak rest-UV.  The AGN would permeate the dust obscuration 
in the radio, however one would expect a much higher 24\um\ flux density due to a high
dust content \citep[by at least an order of magnitude][]{alexander08a}.

\section{Conclusions}

We have presented multiwavelength observations of two high redshift galaxies, arguing 
that they are highly evolved giant elliptical galaxies with low-luminosity, beamed 
AGN.  Their near-IR morphologies (NICMOS H-band) show giant elliptical characteristics 
typical of an old stellar population whose stellar bump is well characterized in the 
IRAC filters.  The physical sizes of both galaxies are 2-5 times larger than most evolved 
galaxies at the same redshift, and similarly the stellar masses are large, around 
$\sim$4$\times$10$^{11}$\,M$_{\sun}$. By the M$_{\star}$-M$_{BH}$ relation, we infer black 
hole masses $\sim$3$\times$10$^{8}$\,M$_{\sun}$. The lack of strong AGN suggests accretion 
at very low rates.  Without X-Ray detections and without strong power law continua in 
the rest-UV, we assume they are low-luminosity AGN whose radio emission must be beamed 
by a factor of a few.  We offer discussions of other possible scenarios, but it is most 
likely that we are observing evolved galaxies with AGN that have nearly exhausted their 
fuel source.  While this type of highly evolved galaxy is rare at high-z, the supplemental 
knowledge that galaxy over-densities have been found near $z=1.99$ in the HDF reinforces, 
if only slightly, the notion that galaxy evolution occurs more rapidly in cluster environments.

These galaxies were originally classified as radio-selected star formers, likely with 
hotter dust temperatures than their SMG counterparts because of their 850\um\ non-detections.  
Only their compact MERLIN radio morphologies and weak 24\um\ flux set them apart from high-SFR 
galaxies.  A full analysis of the SFRG population is in Casey et al. (2008), in preparation.  
The planned eMERLIN deep pointings at 1.4GHz and 5GHz in GOODS-N will push detections to 
$\sim$0.5\uJy\ rms, measuring radio spectral indices and spatially separating emission from 
AGN and star formation.

So far these galaxies have gone unnoticed in an otherwise 
well studied deep survey field.  The potential to study more galaxies exhibiting the same 
properties is possible, as many have probably gone un-noted in studies due to incredibly faint 
X-ray and optical fluxes.  Although both objects were measured at 1.4\,GHz with the VLA, it 
was the MERLIN high resolution coverage which distinguished the extreme radio luminosities 
as emanating from the extremely compact galaxy cores.  If HDF147 and HDF130 contain beamed 
low-luminosity AGN as we propose, they provide a testbed for understanding evolved and poorly 
accreting systems at high redshift.  

\section*{ACKNOWLEDGMENTS}

We thank the anonymous referee for insightful comments and suggestions which helped improve the paper.
We thank Chris Simpson for his advice on the AGN components of HDF147 and 
HDF130, Andy Fabian for his input on the extended inverse Compton emission associated with 
HDF130, and Bret Lehmer for his input on the X-Ray emission from hot gas in both galaxies. CMC
thanks the Gates-Cambridge Trust for support, and DMA thanks the Royal Society.


\label{lastpage}
\end{document}